\input{aipcheck}

\documentclass[
    ,final            
  ]
  {aipproc}

\layoutstyle{8x11single}

\begin{document}

\title{SUSY Breaking from Exotic U(1)\footnote{
Talk given at the International Workshop on Grand Unified Theories
(GUT2012) held at Yukawa Institute for Theoretical Physics, Kyoto, Japan.}}

\classification{11.10.Kk, 12.10.-g, 12.60.Jv}
\keywords      {Supersymmetric Model, Orbifold, supersymmtry breaking, Higher-dimensional gauge theory}

\author{Yoshiharu Kawamura\footnote{
E-mail:haru@azusa.shinshu-u.ac.jp}}{
  address={Department of Physics, Shinshu University, Matsumoto 390-8621, Japan}
}



\begin{abstract}
We propose a mechanism that the soft supersymmetry breaking masses
can be induced from the dynamical rearrangement of local $U(1)$ symmetry
in a five-dimensional model.
The $U(1)$ symmetry possesses several extraordinary features.
The eigenstates of $U(1)$ do not equal to those of boundary conditions.
The $U(1)$ charge of standard model particles does not equal to that of superpartners.
A large $U(1)$ charge hierarchy among superpartners and a standard model gauge singlet is necessary 
in order to obtain the masses of $O(1)$TeV.
\end{abstract}

\maketitle


\section{Introduction}

Our goal is to derive soft supersymmetry (SUSY) breaking masses of $O(1)$ TeV
from the dynamical rearrangement of gauge symmetry\cite{DynRe}, 
which is a part of Hosotani mechanism\cite{H},
or throught the effective potential induced by radiative corrections.
At first sight, it seems to be impossible because any unbroken global SUSY cannot 
induce non-vanishing effective potential.
Here we would like to change the point of view and
answer the question what and how our ordinary assumptions should be modified or relaxed
in order to derive wanted terms from the dynamical rearrangement.
On the way, we will encounter triple difficulties
and introduce $U(1)$ gauge symmetries with very strange properties
to overcome them.

What we want to know is the physics around the TeV scale beyond the Standard Model (SM).
The standard tactic is to attack the problems in the SM.
Our related one is the gauge hierarchy problem\cite{G,W}
or naturalness problem\cite{tH,V}.
The problem is that
{\it unnatural fine-tuning is required to obtain the Higgs mass of order weak scale 
or to stabilize the weak scale.}
If nature does not require the fine-tuning for the Higgs mass,
we have the idea that new physics might exist around the TeV scale and/or
a high-energy physics might have little to do with the SM.
It means that some symmetry suppresses or cancels the effects of interactions,
or the coupling between a high-energy physics and the SM is extremely weak.

A powerful candidate to solve the problem is SUSY
because unwanted divergences can be canceled out by the SUSY.
When SUSY is broken softly, logarithmic divergence appears but harmless
if the mass difference is less than of order TeV scale.
Then if nature takes advantage of SUSY to solve the problem,
SUSY might be broken softly around the TeV scale 
and superpartners might appear around the TeV scale.
Based on the SUSY, people usually construct models according to the scanario
that {\it a high-energy physics is described by a quantum field theory (QFT) respecting SUSY,
the SUSY is spontaneously broken in some hidden sector, 
and soft SUSY breaking terms are induced in our visible sector 
by the mediation of some messengers}\cite{N}.

Now it's time to explain the outline of our exotic scenario\cite{K&M}.
Our scenario is that
{\it SUSY is explicitely broken, at some high-energy scale, 
in the presence of extra gauge symmetries in the bulk,
but boundary conditions (BCs) of fields respect $N=1$ SUSY on our brane and
soft SUSY breaking terms are induced from the dynamical rearrangement
of extra gauge symmetries}.
At this stage, the following question arises.
In the presence of explicit SUSY breaking interactions,
the gauge hierarchy problem and the naturalness problem revisit or not?
If the magnitude of corrections were small enough, unnatural fine-tuning would not be required.
For the time being,
we assume that explicit SUSY breaking interactions are extremely weak
without specifying the origin of such breaking terms.

The content of this paper is as follows.
In the next section, the background including preceding works is reviewed 
and basic ingredients of our scenario are listed.
In section 3, a model is presented to illustrate our idea.
Conclusions are given in section 4.

\section{A scenario}

Let us explain the background and our scenario in order.
That is, the relevant preceding study 
for the origin of soft SUSY breaking terms from extra dimensions
and basic ingredients of our scenario are presented.

First relevant preceding work is the Scherk-Schwarz mechanism\cite{S&S}.
In this mechanism, SUSY breaking terms originate from 
the different BCs between fields and those superpartners.
For example, by imposing the following BCs for five-dimensional (5D) scalar field $\phi(x,y)$ 
and its superpartner $\psi(x,y)$:
\begin{eqnarray}
\phi(x, y+2\pi R) = e^{2\pi i \alpha} \phi(x, y)~, ~~
\psi(x, y+2\pi R) = \psi(x, y)~, 
\label{BC-SS}
\end{eqnarray}
the massless mode $\phi^{(0)}(x)$ of scalar field acquires the mass $\alpha /R$
after compactification, and then SUSY is broken down.
Here $\alpha$ is a constant phase 
and $R$ is a radius of the circle $S^1$.
In order to obtain the mass of $O(1)$TeV,
the magnitude of $\alpha$ should take $O(R/{\mbox{TeV}}^{-1})$.
If the extra dimension had a very small size such that $R=O(1/10^{16})\mbox{GeV}^{-1}$, 
$\alpha$ should be tiny such as $O(10^{-13})$.

The Scherk-Schwarz mechanism has been applied to 
the minimal SUSY extension of SM (MSSM) 
on the 5D space-time $M^4 \times (S^1/Z_2)$\cite{P&Q,BH&N}.
Here $M^4$ is the four-dimensional (4D) Minkowski space, and $S^1/Z_2$ is one-dimensional (1D) orbifold,
which is obtained by dividing $S^1$ with the $Z_2$ transformation, $y \to -y$.
The $N=1$ SUSY in 5D theory is regarded as $N=2$ SUSY in 4D language.
The gauginos consist of doublets under $SU(2)_R$.
Here $R$ means a $R$ symmetry.
By imposing the following BCs for gauginos:
\begin{eqnarray}
\left(
\begin{array}{c}
\lambda^{1} \\ 
\lambda^{2}
\end{array}
\right)(x,-y)= \gamma_5 
\left(
\begin{array}{c}
-\lambda^{1} \\ 
\lambda^{2}
\end{array}
\right)(x,y)~,~~
\left(
\begin{array}{c}
\lambda^{1} \\ 
\lambda^{2}
\end{array}
\right)(x,y+2\pi R)= e^{2\pi i \alpha_{\lambda}\tau_2} 
\left(
\begin{array}{c}
\lambda^{1} \\ 
\lambda^{2}
\end{array}
\right)(x,y)~,
\label{BC-gaugino2}
\end{eqnarray}
the SUSY is partially broken down to $N=1$ in 4D language
and the $N=1$ SUSY is softly broken down with the gaugino mass
proportional to the constant phase $\alpha_{\lambda}$.
Here $\tau_2$ is the second component of Pauli matrix.

The next one is the dynamical rearrangement\cite{DynRe}.
In this mechanism, the physical symmetry and spectra are obtained after the determination of
vacuum state fixed by the Wilson line phases 
as a minimum of the effective potential $V_{\rm eff}$.
The following two pictures are gauge equivalent.
One is a system with the periodic BCs for matter fields and the non-vanishing vacuum expectation values (VEVs) 
for extra-dimensional components of gauge fields.
The other one is that with the twisted BCs and the vanishing VEV.
In the case with 5D model, it is summarized schematically as
\begin{eqnarray}
\{ \phi(x, y+2\pi R) = \phi(x, y)~,~~
\langle A_y \rangle = {2\pi \alpha}/{R}\} ~~~ \sim ~~~
\{ \phi(x, y+2\pi R) = e^{2\pi i \alpha} \phi(x, y)~,~~
\langle A_y \rangle = 0\}~,
\label{BC-eq}
\end{eqnarray}
where $A_y$ is the extra-dimensional component of gauge field and $\sim$ stands for the gauge equivalence.
Through the mechanism, the gauge symmetry can be dynamically broken down or restored.

{}From the above observations, the following question arises naturally.
{\it Is it possible to break SUSY from the dynamical rearrangement?
Or does the Scherk-Schwarz mechanism work dynamically?}

If we try to answer affirmatively, we encounter two major barriers suddenly.
One is on SUSY.
Unbroken SUSY usually leads to the vanishing effective potential 
and so the Hosotani mechanism does not work to generate terms we want.
The other one is on broken charge (and SUSY).
To generate soft SUSY breaking masses, 
relevant broken charge of SM particles (except Higgs bosons) should vanish,
but those of superpartners (except Higgsinos) should not vanish.
That is, the SM particles and their superpartners do not have 
a common quantum number for some gauge symmetry.
A possible candidate for relevant broken charge is $SU(2)_R$.
Hence we need a theoretical framework with a local $SU(2)_R$ symmetry.
Furthermore a theory should have a local SUSY
because $SU(2)_R$ is not orthogonal to SUSY.

As a framework with a local $SU(2)_R$ symmetry, 
5D supergravity (SUGRA) is known\cite{K&O,Z},
and the breakdown of SUSY from the Hosotani mechanism has been studied
in this framework\cite{G&Q,GQ&R}.
{}From the 5D SUGRA including $S^1/Z_2$ as the extra space,
the following effective potential is obtained
\begin{eqnarray}
V_{\rm eff} = \frac{3}{32 \pi^6 R^4} (2+N_V-N_H) \sum_{n=1}^{\infty} \frac{1}{n^5}\cos(2\pi n \alpha)~,
\label{VeffSUGRA}
\end{eqnarray}
where $N_V$ and $N_H$ are numbers of vectormultipletes and hypermultiplets in the bulk, respectively.
{}From (\ref{VeffSUGRA}),
only $\alpha= 0$ or $1/2$ can be obtained in the absence of other SUSY breaking sources.
Then it leads to unbroken SUSY or large SUSY breaking if $R$ is very small.
Here, the third barrier appears, that is,
how we should obtain the SUSY breaking masses with an appropriate size.
This is a tough barrier because the result (\ref{VeffSUGRA}) is robust 
from the fact that the effective potential is relevant to the numbers of particle contents
and the normalization of $SU(2)_R$ charges is fixed from the group theoretical reason.
In this way, we arrive the no go theorem that
{\it soft SUSY breaking parameters of $O(1)$TeV cannot be obtained via the Hosotani mechanism
from any SUSY QFT without SUSY breaking sources and with flat small extra dimensions}.

We would like to escape the no go theorem by modifying or relaxing assumptions.
In fact, we find that {\it there is a chance if SUSY were broken in the bulk in the presence of exotic $U(1)$}.
Here Why $U(1)$? What does the exotic $U(1)$ mean?
And how do SUSY breaking terms come from?
Those are questions we will address.

Next we explain the basic ingredients of our scenario.
They consist of six parts. 
\begin{itemize}
\item[(1)] The space-time is made from a product of $M^4$ and the extra space $O$. 
Our 4D world is a brane or boundary in the bulk.

\item[(2)] The relevant gauge group is $G_{\rm SM} \times G'$.
Here $G_{\rm SM}$ is the SM gauge group $G_{\rm SM} = SU(3)_C \times SU(2)_L \times U(1)_Y$
and $G'$ is an exotic gauge group.
Gauge multiplets live in the bulk.

\item[(3)] The same number of bosonic fields and fermionic ones exits, 
e.g., as a remnant of SUSY at a higher energy scale beyond our starting QFT.
The corresponding partners have a same quantum number under $G_{\rm SM}$,
{\it but a different quantum number under $G'$}.\footnote{
We do not specify the origin of exotic gauge symmetries.
As a conjecture, bulk fields with $G'$ quantum numbers might 
be solitonic objects originating from unknown non-perturbative dynamics
on the formation of space-time structure in a more fundamental theory.
Or they might be survivers from SUSY multiplets after decoupling some partners.}
Hence {\it the SUSY is manifestly broken in the bulk at a turn of the switch of $G'$.}
Then the SUSY breaking effects must mediate on the brane.
To avoid the gauge hierarchy problem and the naturalness problem, 
we assume that the coupling between the SUSY breaking source and the SM sector
is extremely weak with a tiny gauge coupling and/or a tiny charge.

\item[(4)] The $G'$ is broken down to its subgroup $H'$ on our brane
by suitable BCs relating the extra dimension, which respect the SUSY.
Hence {\it the $N=1$ SUSY can be realized in the low-energy spectra on our brane} at the tree level.
We assume that all fields are singlets (or fields and their would-be superpartners have a common quantum number) 
under $H'$.

\item[(5)] The MSSM fields come from zero modes of bulk fields and our brane fields.
Physics can be described as the MSSM without soft SUSY breaking terms on our brane at the tree level.

\item[(6)] The dynamical rearrangement of $G'/H'$ occurs and soft SUSY breaking masses are induced.
Those masses are proportional to the VEVs of extra-dimensional components of broken gauge bosons
and broken charges, that is, generators of $G'/H'$.
To obtain $m_{\tiny{\mbox{SUSY}}}$ of $O(1)$TeV,
tiny charges are required and so $U(1)$ symmetry is suitable.
(This is the answer for the question why $U(1)$.)
\end{itemize}

\section{A Model}

Let us illustrate our scenario in a more abstract form.
The space-time is assumed to be $M^4 \times (S^1/Z_2)$.
We introduce the exotic $U(1)$ gauge field $A_M^{(-)}(x,y)$ $(M = 0, 1, 2, 3, 5)$ 
with the following BCs:
\begin{eqnarray}
\hspace{-0.7cm} &~& A_M^{(-)}(x,y+2\pi R)=A_M^{(-)}(x,y)~,~~
\label{Apm-BC1}\\
\hspace{-0.7cm} &~& A_{\mu}^{(-)}(x,-y)=- A_{\mu}^{(-)}(x,y)~,~~ A_{5}^{(-)}(x,-y)= A_{5}^{(-)}(x,y)~,
\label{Apm-BC2}\\
\hspace{-0.7cm} &~& A_{\mu}^{(-)}(x, 2\pi R-y)=- A_{\mu}^{(-)}(x,y)~,~~ A_{5}^{(-)}(x,2\pi R-y)= A_{5}^{(-)}(x,y)~.
\label{Apm-BC3}
\end{eqnarray}
This $U(1)$ is broken down by the above BCs 
because the massless mode of $A_{\mu}^{(-)}$ $(\mu = 0, 1, 2, 3)$ is absent.
On the other hand, the massless mode of $A_{5}^{(-)}$ survives and becomes a dynamical one,
which plays a central role on the dynamical rearrangement.

Let us explain a way to assign $Z_2$ even parity for $A_{5}^{(-)}$.
This method is a variant given in Ref. \cite{KT&Y}.
We introduce a doublet $\phi_k$ ($k=1,2$) under the $Z_2$ reflection
whose BC is given by,
\begin{eqnarray}
\phi_k(x,y+2\pi R)=\eta_{0}\phi_k(x,y)~,~~ \phi_1(x,-y)=\eta_{1}\phi_2(x,y)~,~~
\phi_1(x,2\pi R-y)=\eta_{2}\phi_2(x,y)~,
\label{phi-BC}
\end{eqnarray}
where $\eta_{1}$ and $\eta_{2}$ are intrinsic $Z_2$ parities
whose vaules are $+1$ or $-1$, and $\eta_{0} = \eta_{1}\eta_{2}$.
We can construct the $Z_2$ invariant Lagrangian density.
For example, the extra-coordinate part of kinetic term is given by
\begin{eqnarray}
\left|\left(
\begin{array}{cc}
\partial_5 +iq_-A_5^{(-)}& 0 \\
0 & \partial_5 -iq_-A_5^{(-)}
\end{array}
\right)
\!\!
\left(
\begin{array}{c}
\phi_1 \\ 
\phi_2
\end{array}
\right)\right|^2~,
\label{Phi-Dy}
\end{eqnarray}
where we omit the SM gauge bosons irrelevant of our discussion
and the gauge coupling of $U(1)^{(-)}$ to avoid a complication.
{}From (\ref{Phi-Dy}), the $Z_2$ doublet is the eigenstate of $U(1)$.
On the other hand, the eigenstates of BCs are constructed from linear combinations such that
$\phi_{\pm} \equiv (\phi_1 \pm \phi_2)/\sqrt{2}$
and they obey the following BCs:
\begin{eqnarray}
\phi_{\pm}(x,y+2\pi R)=\eta_{0}\phi_{\pm}(x,y)~,~~
\phi_{\pm}(x,-y)= \pm \eta_{1}\phi_{\pm}(x,y)~,~~
\phi_{\pm}(x,2\pi R-y)=\pm \eta_{2}\phi_{\pm}(x,y)~.
\label{phipm-BC}
\end{eqnarray}
Using them, (\ref{Phi-Dy}) is rewritten by
\begin{eqnarray}
\left|
\left(
\begin{array}{cc}
\partial_5  & iq_{-}A_5^{(-)} \\
iq_{-}A_5^{(-)} & \partial_5 
\end{array}
\right)
\!\!
\left(
\begin{array}{c}
\phi_{+} \\ 
\phi_{-}
\end{array}
\right)\right|^2 = q_{-}^2 \left( A_5^{(-)} \right)^2 \left| \phi_+^{(0)} \right|^2 + \cdots~,
\label{Phipm-Dy}
\end{eqnarray}
where $\phi_+^{(0)}$ is a zero mode.
We find an $SU(2)$-like structure in the first expression of (\ref{Phipm-Dy}).
In this way, mass terms can be obtained if $A_{5}^{(-)}$ acquires the non-vanishing VEV.
The question left behind is how to derive $\langle A_{5}^{(-)} \rangle$ with a suitable size.

The BCs of MSSM fields in the 5D space-time are given by
\begin{eqnarray}
&~& A_{M}(x,y+2\pi R)= A_{M}(x,y)~,~~ \Sigma(x,y+2\pi R)= \Sigma(x,y)~,~~ 
\lambda^{i}(x,y+2\pi R)= \lambda^{i}(x,y)~,
\label{Gauge-BC1}\\ 
&~& A_{\mu}(x,-y)= A_{\mu}(x,y)~,~~ 
\left(
\begin{array}{c}
A_{5} \\ 
\Sigma
\end{array}
\right)(x,-y)
= -\left(
\begin{array}{c}
A_{5} \\ 
\Sigma
\end{array}
\right)(x,y)~,~~ 
\nonumber \\
&~& \lambda^{1}(x,-y)= -\gamma_5 \lambda^{1}(x,y)~,~~ \lambda^{2}(x,-y)= \gamma_5 \lambda^{2}(x,y)~,
\label{Gauge-BC2}\\
&~& A_{\mu}(x,2\pi R-y)= A_{\mu}(x,y)~,~~ 
\left(
\begin{array}{c}
A_{5} \\ 
\Sigma
\end{array}
\right)(x,2\pi R-y)
= -\left(
\begin{array}{c}
A_{5} \\ 
\Sigma
\end{array}
\right)(x,y)~,~~ 
\nonumber \\
&~& \lambda^{1}(x,2\pi R-y)= -\gamma_5\lambda^{1}(x,y)~,~~ \lambda^{2}(x,2\pi R-y)= \gamma_5\lambda^{2}(x,y)~,
\label{Gauge-BC3}\\
&~& \psi^i(x,y+2\pi R)= \psi^i(x,y)~,
\left(
\begin{array}{c}
\phi^i \\ 
\phi^{ci\dagger}
\end{array}
\right)\!\!(x,y+2\pi R)= 
\left(
\begin{array}{c}
\phi^i \\ 
\phi^{ci\dagger}
\end{array}
\right)\!\!(x,y)~,~ 
\label{Matter-BC1}\\ 
&~& \psi^i(x, -y)= -\gamma_5 \psi^i(x,y)~,
\left(
\begin{array}{c}
\phi^i \\ 
\phi^{ci\dagger}
\end{array}
\right)\!\!(x, -y)= 
\left(
\begin{array}{c}
\phi^i \\ 
-\phi^{ci\dagger}
\end{array}
\right)\!\!(x,y)~,
\label{Matter-BC2}\\
&~& \psi^i(x, 2\pi R-y)= -\gamma_5 \psi^i(x,y)~,
\left(
\begin{array}{c}
\phi^i \\ 
\phi^{ci\dagger}
\end{array}
\right)\!\!(x, 2\pi R-y)= 
\left(
\begin{array}{c}
\phi^i \\ 
-\phi^{ci\dagger}
\end{array}
\right)\!\!(x,y)~,
\label{Matter-BC3}
\end{eqnarray}
where $A_M$ is the 5D SM gauge bosons, $\Sigma$ is a real scalar field, $(\lambda^1, \lambda^2)$
are gauginos, $\psi^i$ are fermions represented by four-component spinors 
and $(\phi^i, \phi^{ci\dagger})$ are complex scalar fields.
The index indicating the SM gauge group or generators is suppressed
and $i$ represents particle species.
Here, those fields are given by the eigenstates of BCs, e.g., 
$\phi^i$ and $\phi^{ci\dagger}$ are regarded as $\phi_+$ and $\phi_-$, respectively,
with $\eta_{0} = \eta_{1} = \eta_{2} = 1$.
We assume that sfermions, gauginos and Higgs bosons have the non-vanishing $U(1)$ charge $q_{s}$
for the eigenstates of $U(1)$ gauge symmetry,
$q_a$ and $q_h$, but those partners do not.
Then we obtain the following effective potential for 5D MSSM particles:
\begin{eqnarray}
V_{\rm eff}^{\mbox{\tiny{MSSM}}}[\beta] 
= -4C \sum_{s} \sum_{n=1}^{\infty}{1 \over n^5}\cos{\left[2\pi n q_{s}\beta\right]}
+ 4C \sum_{a} \sum_{n=1}^{\infty}{1 \over n^5}\cos{\left[2\pi n q_{a}\beta\right]} 
-8C \sum_{h} \sum_{n=1}^{\infty}{1 \over n^5}\cos{\left[2\pi n q_{h}\beta\right]} ~,
\label{MSSM-Veff}
\end{eqnarray}
where $C = 3/(128 \pi^6 R^4)$ and $\beta \equiv \langle A_{5}^{(-)} \rangle R$.
We find that the effective potential is minimized at $\beta = 0$ 
if $q_s$s have the same magnitude of $q_a$ and $q_h$.
Then SUSY is unbroken.

By changing the BCs with $\eta_0 =1$ into those with $\eta_0 =-1$
for sfermions, 
the following effective potential is obtained
\begin{eqnarray}
V_{\rm eff}^{\mbox{\tiny{MSSM}}}[\beta] 
= -4C \sum_{i} \sum_{n=1}^{\infty}{1 \over n^5}\cos{\left[2\pi n \left(\frac{1}{2} + q_{s}\beta\right)\right]}
+ 4C \sum_{a} \sum_{n=1}^{\infty}{1 \over n^5}\cos{\left[2\pi n q_{a}\beta\right]} 
-8C \sum_{h} \sum_{n=1}^{\infty}{1 \over n^5}\cos{\left[2\pi n q_{h}\beta\right]} ~.
\label{MSSM-Veff-SUSYbr}
\end{eqnarray}
We find that the effective potential is minimized at $q \beta = 1/2$
where we take a common value $q$ for $q_s$ for simplicity.
Then SUSY is broken down, but much bigger soft SUSY breaking masses are obtained if $R$ is very small.
This result is similar to the case of 5D SUGRA.

In order to construct a realistic model,
we introduce a sector with a SM gauge singlet $\Phi$ whose $U(1)$ charge is $q_{\Phi}$,
which leads to the following effective potential:
\begin{eqnarray}
V_{\rm eff}^{\Phi}[\beta] 
 = -4C \sum_{n=1}^{\infty}{1 \over n^5}\cos{\left[2\pi n \left(\frac{1}{2} + q_{\Phi}\beta\right)\right]}~.
\label{Psi-Veff}
\end{eqnarray}
If $q_{\Phi}$ is much bigger than others, 
this potential $V_{\rm eff}^{\Phi}[\beta]$ dominates over the MSSM one
and the minimum is given by $q_{\Phi} \beta = 1/2$ and SUSY is broken down.
Then soft SUSY breaking masses of $O(1)$TeV can be derived
if $q_{\rm sp}$s are tiny compared with $q_{\Phi}$.
Here $q_{s}$, $q_{a}$ and $q_{h}$ are denoted as $q_{\rm sp}$ as a whole.
Let us explain it furthermore.
The magnitude of $\langle q_{\Phi} A_5^{(-)} \rangle$ is estimated as $1/(2R)$
and that of soft SUSY breaking masses is estimated as $q_{\rm sp}/(2q_{\Phi}R)$.
Even if $1/R$ is $O(10^{16})$GeV, the masses of $O(1)$TeV are obtained
with a large hierarchy such that $q_{\rm sp}/q_{\Phi} =O(10^{-13})$
or a very tiny charge $q_{\rm sp}$ unless the magnitude of $q_{\Phi}$ is big.
Hence we find that
{\it a large charge hierarchy of $q_{\rm sp}/q_{\Phi} = O(R/\mbox{TeV}^{-1})$ is necessary 
in order to obtain the soft SUSY breaking masses of $O(1)$TeV.}
It is a difficult problem whether such a large charge hierarchy is derived naturally.
This is one of problems in our scenario.

The $\mu$ parameter can be induced 
by the dynamical rearrangement of another $U(1)$ gauge symmetry\cite{K&M}.\footnote{
In Ref. \cite{K&M2}, the generation of $\mu$ parameter from the dynamical rearrangement 
of local $U(1)$ symmetry in a five-dimensional model has been studied
under the assumption that SUSY is broken by the Scherk-Schwarz mechanism.
}

\section{Conclusion}

We have proposed a scenario with an illustrating model.
Our scenario is that
SUSY is explicitely broken, at some high-energy scale, 
in the presence of extra gauge symmetries in the bulk,
but BCs of fields respect $N=1$ SUSY on our brane and
soft SUSY breaking terms are induced from the dynamical rearrangement
of extra gauge symmetries.

The big problem in our scenario is the origin of exotic $U(1)$ symmetry.
The exotics come from the following three bizarre features.
First one is that the eigenstates of $U(1)$ symmetry do not equal to those of BCs
and it can make $Z_2$ parities of $A_5^{(-)}$ even.
Then it turns out to be a seed of dynamical rearrangement.
The second one is that the $U(1)$ charge of SM particles does not equal to that of superpartners
and it turns out to be a seed of SUSY breaking.
The last one is that there is a large charge hierarchy of $q_{\rm sp}/q_{\Phi} = O(R/\mbox{TeV}^{-1})$ 
and it turns out to be a seed of TeV scale.

Another big problem is how to formulate our scenario in the framework of SUGRA.
In the presence of explicit breaking terms for a local symmetry,
the theory can, in general, fall into the inconsistency
such as the breakdown of unitarity and so on.

\begin{theacknowledgments}
I would like to thank T.~Kugo, C.S.~Lim, N.~Maru, Y.~Sakamura and T.~Yamashita for useful discussions.
This work was supported in part by scientific grants from the Ministry of Education, Culture,
Sports, Science and Technology under Grant Nos.~22540272 and 21244036.
\end{theacknowledgments}



\bibliographystyle{aipproc}   

\bibliography{sample}

\begin{thebibliography}{9}

\bibitem{DynRe} N.~Haba, M.~Harada, Y.~Hosotani and Y.~Kawamura, \emph{Nucl. Phys.} \textbf{B657}, 169 (2003);
\emph{Prog. Theor. Phys.} \textbf{111}, 265 (2004).

\bibitem{H} Y.~Hosotani, \emph{Phys. Lett.} \textbf{B126}, 309 (1983); \emph{Ann. Phys.} \textbf{190}, 233 (1989).

\bibitem{G} E.~Gildener, \emph{Phys. Rev.} \textbf{D14}, 1667 (1976).

\bibitem{W} S.~Weinberg, \emph{Phys. Lett.} \textbf{82B}, 387 (1979).

\bibitem{tH} G.~'t~Hooft, \emph{Proc. NATO Advanced Study Institute on Recent Developments in
Gauge theories}, 135 (1979).

\bibitem{V} M.~J.~G.~Veltman, \emph{Acta. Phys. Pol.} \textbf{B12}, 437 (1981).

\bibitem{N} For a review, see H.~P.~Nilles, \emph{Phys. Rep.} \textbf{110}, 1 (1984).

\bibitem{K&M} Y.~Kawamura and T.~Miura, \emph{Int. J. Mod. Phys.} \textbf{26}, 4405 (2011). 

\bibitem{S&S} J.~Scherk and J.~H.~Schwarz, \emph{Phys. Lett.} \textbf{82B}, 60 (1979);
\emph{Nucl. Phys.} \textbf{B153}, 61 (1979).

\bibitem{P&Q} A.~Pomarol and M.~Quir\'os, \emph{Phys. Lett.} \textbf{B438}, 255 (1998).

\bibitem{BH&N} R.~Barbieri, L.~Hall and Y.~Nomura, \emph{Phys. Rev.} \textbf{D66}, 045025 (2002).

\bibitem{K&O} T.~Kugo and K.~Ohashi, \emph{Prog. Theor. Phys.} \textbf{104}, 835 (2000); \textbf{105}, 323 (2001).

\bibitem{Z} M.~Zucker, \emph{Nucl. Phys.} \textbf{B570}, 267 (2000);
\emph{JHEP} \textbf{0008}, 016 (2000).

\bibitem{G&Q} G.~v.~Gersdorff and M.~Quir\'os, \emph{Phys. Rev.} \textbf{D65}, 064016 (2002).

\bibitem{GQ&R} G.~v.~Gersdorff, M.~Quir\'os and A.~Riotto, \emph{Nucl. Phys.} \textbf{B634}, 90 (2002).

\bibitem{KT&Y} K.~Kojima, K.~Takenaga and T.~Yamashita, \emph{Phys. Rev.} \textbf{D84}, 051701 (2011).

\bibitem{K&M2} Y.~Kawamura and T.~Miura, \emph{Int. J. Mod. Phys.} \textbf{27}, 1250023 (2012). 

\end{thebibliography}

\IfFileExists{\jobname.bbl}{}
 {\typeout{}
  \typeout{******************************************}
  \typeout{** Please run "bibtex \jobname" to optain}
  \typeout{** the bibliography and then re-run LaTeX}
  \typeout{** twice to fix the references!}
  \typeout{******************************************}
  \typeout{}
 }

\end{document}